\documentclass[11pt,a4paper]{scrartcl}

\usepackage{ILD}

\usepackage{xpatch}
\makeatletter
\xpatchcmd\@HepConStyle
 {\edef\@upcode{\updefault}}
 {\ifdefined\shapedefault\edef\@upcode{\shapedefault}\else\edef\@upcode{\updefault}\fi}
 {}{}
\makeatother

\usepackage[symbol]{footmisc}
\usepackage{feynmf}
\usepackage{multirow}
\usepackage{textpos}
\usepackage{heppennames2}
\usepackage{LCVision_definitions}

\usepackage{siunitx}
\usepackage{lineno}

\usepackage{booktabs,multirow,array}
\usepackage{amssymb}

\title{Higgs Self-Coupling Measurement \\at a Linear Collider at 550 GeV}

\ildproc{phys}{2026}{003}

\date{\today}

\addauthor{Mikael Berggren}{\institute{1}}
\addauthor{Bryan Bliewert}{\institute{1}\institute{2}}
\addauthor{Jenny List}{\institute{1}}
\addauthor{Dimitris Ntounis}{\institute{3}}
\addauthor{Taikan Suehara}{\institute{4}}
\addauthor{Junping Tian}{\institute{4}}
\addauthor{Julie Munch Torndal}{\institute{1}\institute{2}}
\addauthor{Caterina Vernieri}{\institute{3}}
\addinstitute{1}{Deutsches Elektronen-Synchrotron DESY, Germany}
\addinstitute{2}{Department of Physics, Universit\"at Hamburg, Germany}
\addinstitute{3}{SLAC National Accelerator Laboratory, United States}
\addinstitute{4}{International Center for Elementary Particle Physics (ICEPP), The University of Tokyo, Japan}

\abstract{The Higgs mechanism is essential for the success of the Standard Model (SM) and can be experimentally verified with the determination of the Higgs self-coupling. As the simplest model of a Higgs potential, the SM provides a clear prediction of the Higgs self-coupling in terms of the Higgs boson mass and the vacuum expectation value. Any deviations would indicate physics beyond the SM and help guide extended Higgs models. At large enough centre-of-mass energies, double-Higgs production provides tree-level sensitivity to the trilinear Higgs self-coupling. At 550\,GeV the leading production mode in $\Pep\Pem$ comes from di-Higgs strahlung with a small contribution from $\PW\PW$-fusion. The most up-to-date ILD projections are extrapolated based on a full simulation analysis from 2014 by incorporating expected improvements in flavour tagging and kinematic reconstruction for event selection, and are presented in this contribution together with the ongoing re-analysis using fast SGV (Simulation a Grande Vitesse) simulations of the ILD detector concept on a full SM background including the aforementioned state-of-the-art reconstruction and analysis tools.\\\\

Talk presented at the International Workshop on Future Linear Colliders (LCWS 2025), October 20-24 2025.}

\DeclareTOCStyleEntry[beforeskip=.2cm]{section}{section}

\graphicspath{ {./}{./figures/} }

\begin{document}

\titlepage
\pagenumbering{arabic}\setcounter{page}{1}

\clearpage

\section{Introduction}
\begin{figure}[htbp]
    \centering
    \includegraphics[width=0.5\textwidth]{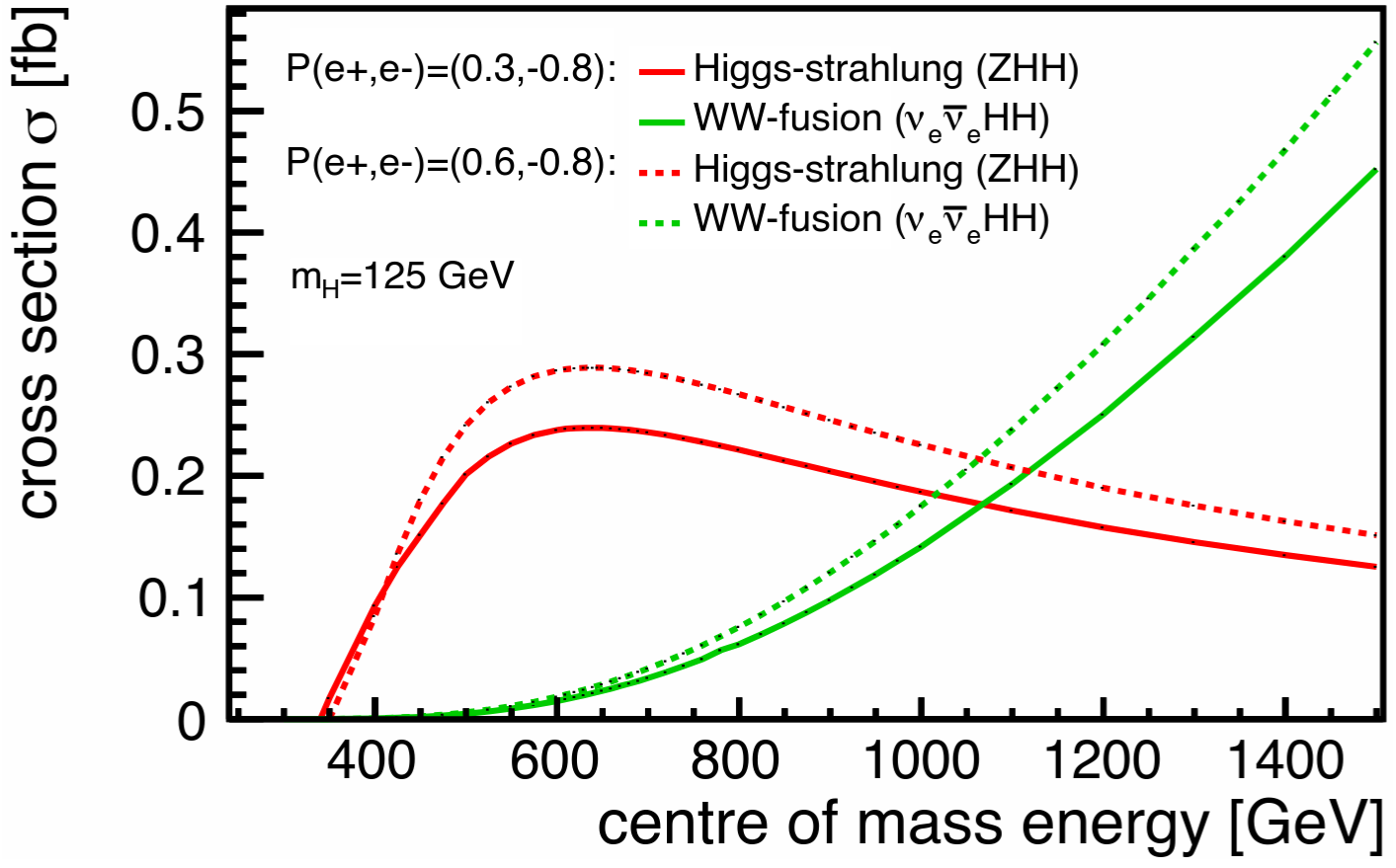}
    \caption{Cross sections of double Higgs-strahlung and $\PW\PW$-fusion as a function of the centre-of-mass energy for two polarisation modes~\cite{Durig:2016jrs}.}
    \label{fig:xsec_claude}
\end{figure}

The trilinear Higgs self-coupling, $\uplambda_{\PH\PH\PH}$, can be accessed at tree-level through double-Higgs production. The Higgs self-coupling measurement is challenging due to the small cross sections of double Higgs production. There are two main production modes, the double Higgs-strahlung with a $\PZ\PH\PH$ final state and $\PW\PW$-fusion with a $\PGn\PAGn\PH\PH$ final state. Their cross sections as a function of centre-of-mass energy can seen be in Fig.~\ref{fig:xsec_claude}, showing that below 1\,TeV the leading contribution comes from $\PZ\PH\PH$ while above 1\,TeV, the $\PGn\PAGn\PH\PH$ contribution becomes dominant. An analysis effort~\cite{Durig:2016jrs, Tian:2013, Kurata:2013} considering the $\PH\PH\to 4\PQb$ and $\PH\PH\to \PQb\PAQb\PW\PW^*$ final states from 2014 showed that linear $\Pep\Pem$ colliders have discovery reach of the di-Higgs production at 500\,GeV where an $8\sigma$ observation of the $\Pep\Pem\to \PZ\PH\PH$ process was projected. But due to interference effects between the $\PZ\PH\PH$ final state diagrams with and without the Higgs self-coupling, this only corresponded to a 27\% precision on the $\uplambda_{SM}$. The interference effects are conveyed through a sensitivity factor, $k$, which translates the cross section $\Delta\sigma_{\PH\PH x}$ to a value of $\uplambda_{\PH\PH\PH}$ via
\begin{equation}
  \label{eq:precision}
  \frac{\Delta\uplambda_{\PH\PH\PH}}{\uplambda_{\PH\PH\PH}}=k\cdot\frac{\Delta\sigma_{\PH\PH x}}{\sigma_{\PH\PH x}}.
\end{equation}

Recently, a new analysis effort on $\PZ\PH\PH$ has been started to update the projections on the Higgs self-coupling, movitated by large advancements in tools for event reconstruction and analysis algorithms, often benefitting from the introduction of machine learning, that we have seen since the 2014 analyses, as well as a revision of running scenarios for linear $\Pep\Pem$ colliders, where the physics programmes for the International Linear Collider (ILC)~\cite{TDR1,TDR2,TDR31,TDR32,TDR4}, the Cool Copper Collider (C$^3$)~\cite{Vernieri:2022fae}, and the Linear Collider Facility (LCF)~\cite{LinearCollider:2025lya} all plan to run at 550\,GeV instead of the previously considered 500\,GeV. Additionally, the LCF also plans for doubling the integrated luminosty to 8\,ab$^{-1}$, compared to 4\,ab$^{-1}$ for the ILC and C$^3$ at 550\,GeV, as well as doubling the beam polarisation from 30\% for the positron source at the ILC and C$^3$ to 60\% for the LCF still with the 80\% electron beam polarisation. The effect of increasing the positron beam polarisation on the cross sections of $\PZ\PH\PH$ and $\PGn\PAGn\PH\PH$ can also be seen in Fig.~\ref{fig:xsec_claude}.

While the current analysis is still ongoing, an update on the projections on the $\uplambda_{\PH\PH\PH}$ determination has recently been published in an ILD note~\cite{ZHHnote} in time to have contributed to the Briefing Book~\cite{briefingbook} for the next update of the European Strategy for Particle Physics. The note briefly summarises the main conclusions of the 2014 analyses, discusses the developments in relevant high-level reconstruction tools, provides first projections on the precision reach in the lepton channel and briefly comments on the neutrino channel, and finally gives the full projections on the Higgs self-coupling based on extrapolations from the 2014 analysis using the relative improvements derived from our advancements in analysis tools, as well as considering the revised running scenarios. This contribution presents a summary of that note, focusing on the improvements in reconstruction tools, the results of the lepton channel to confirm the extrapolation, and finally the results of the extrapolations both for the SM case and beyond the SM.   

\section{Improvements in reconstruction tools}
The $\PZ\PH\PH$ re-analysis has been seen to improve significantly with the advancements both in flavor tagging and in the kinematic reconstruction. Flavor tagging is crucial for the $\PH\PH\to 4\PQb$ channel and has improved significantly since the last analysis with the introduction of advanced machine learning. A decade ago, the tool for flavor tagging, LCFIPlus~\cite{Suehara:2015ura}, was based on BDTs which categorised jets into $\PQb$, $\PQc$ and light jets based on the BDT score. Nowadays, the state-of-the-art tool for flavor tagging, PartT, is based on a particle transformer architecture~\cite{Qu:2022mxj, Tagami:2024gtc} enabling the use of a large amount of jet information. ParT is able to categorise the jets into 3 ($\PQb$, $\PQc$ and light jets), 6 ($\PQb$, $\PQc$, $\PQs$, $\PQu$, $\PQd$, $\Pg$) or 11 ($\PQb$, $\PAQb$, $\PQc$, $\PAQc$, $\PQs$, $\PAQs$, $\PQu$, $\PAQu$, $\PQd$, $\PAQd$, $\Pg$) jet categories. Table~\ref{tab:misrate} shows how the $\PQb$-tag efficiency for various $\PQc$-jet mis-identity rates improves between LCFIPlus and PartT for various configurations of PartT. Comparing the 3-category PartT to LCFIPlus in rows 1-2 shows a 15\%, 20\% and 25\% higher $\PQb$-jet efficiency for the fixed 10\%, 3\% and 1\% $\PQc$-jet mis-identity rates. When extrapolating the Higgs self-coupling projections from the 2014 analysis, a more conservative 10\% higher tagging accuracy has been assumed. The ongoing analysis uses an 11-category version of PartT trained on 800k fast SGV~\cite{Berggren:2012ar} simulation jets. Rows 3-4 show that the PartT performance is comparable between full simulation and SGV. And rows 4-5 show that there is potential for improvement by using a newer version of PartT trained on 8M events.

\begin{table}[htb]
  \centering                                                                                                                                      
  \begin{tabular}{lrrr }
    \hline
    $\PQc$-jet mis-ID rate & 10\% & 3\% & 1\% \\
    \hline
    LCFIPlus, 91\,GeV $\PQq\PAQq$ and fullsim                               & 82\%  & 74\% & 67\% \\
    ParT, 250\,GeV $\PGn\PAGn\PQq\PAQq$ and fullsim, 800k jets, 3-category    & 95\%  & 90\% & 85\% \\
    ParT, 250\,GeV $\PGn\PAGn\PQq\PAQq$ and fullsim, 800k jets, 11-category   & 95\%  & 90\% & 83\% \\
    ParT, 250\,GeV $\PGn\PAGn\PQq\PAQq$ and SGV, 800k jets, 11-category       & 94\%  & 90\% & 87\% \\
    ParT, 250\,GeV $\PGn\PAGn\PQq\PAQq$ and SGV, 8M jets, 11-category  & 95\%  & 92\% & 89\% \\
    \hline
  \end{tabular}
  \caption{The $\PQb$-tag efficiency for three working points of the $\PQc$-background rejection with various configurations of PartT~\cite{ZHHnote}.}
  \label{tab:misrate}
\end{table}

\begin{figure}[htbp]
    \centering
    \begin{subfigure}{.5\textwidth}
        \centering
        \includegraphics[width=0.75\textwidth,page=1]{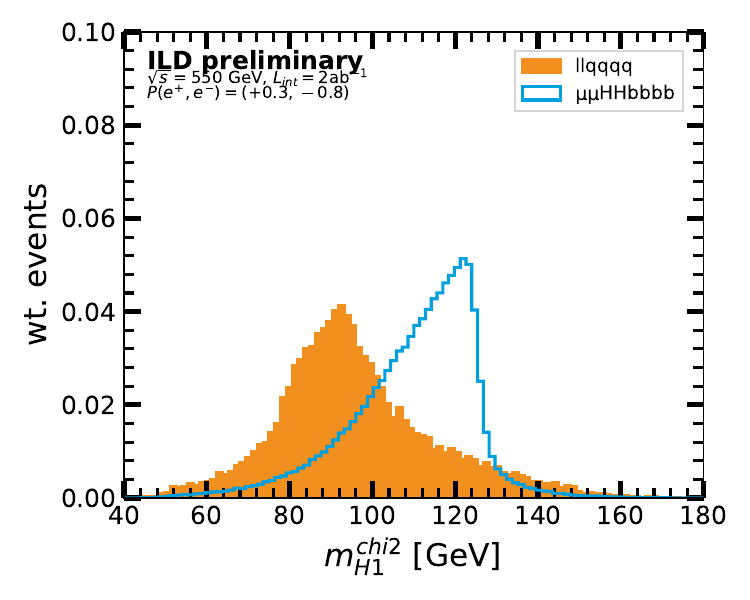}
        \caption{}
        \label{fig:mhllqqqq:mh1chi2}
    \end{subfigure}%
    \begin{subfigure}{.5\textwidth}
        \centering
        \includegraphics[width=0.75\textwidth,page=3]{Bryan_mHi_llqqqq_mumubbbb_250830.pdf}
        \caption{}
        \label{fig:mhllqqqq:mh1kinfit}
    \end{subfigure}%
    \caption{Invariant mass of the ``least Higgs-boson-like'' di-jet mass before (a) and after (b) a 4C-fit for the muon channel with $\PZ\PH\PH\to\PGmp\PGmm\PQb\PAQb\PQb\PAQb$ compared to $\Pl\Pl\PQq\PQq\PQq\PQq$ background events~\cite{ZHHnote}.
    }
    \label{fig:mhllqqqq}
\end{figure}

One of the main limitations of the $\PZ\PH\PH$ analysis is jet-misclustering, degrading the mass resolutions of jets and jet systems as seen in Fig.~2 of~\cite{ZHHnote} which compares the di-jet masses between jet clustering using the Durham algorithm~\cite{Durham} and having ``perfect'' jet clustering by cheating the association of the jet constituents. ``Perfect'' jet clustering has the potential to improve the relative precision on the cross section by up to 40\%~\cite{Durig:2016jrs}.
Unfortunately, no new methods exist that are better than the Durham algorithm but the di-jet mass can be improved in other ways. The well-known initial state in $\Pep\Pem$ colliders can be exploited with kinematic fitting and is used to improve the event reconstruction, jet pairing and hypothesis testing. In the simplest kinematic fit used, a 4C-fit, only 4-momentum conservation is required, introducing four constraints. The fit provides a $\chi^2$ measure and a probability as well as corrections to the fit objects i.e. the jets, leptons, and ISR. Well-estimated errors are crucial for a well-functioning fit which are provided with ErrorFlow~\cite{Radkhorrami:2024phd}, parametrizing the sources of uncertainties for the individual jets from detector resolution as well as particle confusion from the particle flow algorithm. Additionally, a correction of unmeasured neutrinos in a jet is calculated from its kinematics down to a sign-ambiguity, resolved by the 4C-fit~\cite{Radkhorrami:2024phd}. The 4C-fit improves the mass resolution of the di-jets as seen in Fig.~\ref{fig:mhllqqqq}, where the di-jet mass for the ``least Higgs-boson-like'' di-jet in the  muon channel with $\PGmp\PGmm\PQb\PAQb\PQb\PAQb$ events compared to 6-fermion background events with a $\Pl\Pl\PQq\PQq\PQq\PQq$ final state is plotted previous to the 4C-fit in \ref{fig:mhllqqqq:mh1chi2}, where the jet pairing is decided from minimising a simple $\chi^2$-calculation, and posterior to the 4C-fit in \ref{fig:mhllqqqq:mh1kinfit}, where the jet pairing is decided from a subsequent 6C-fit where two additional mass constraints on the two di-jets are imposed on each event. The resolution on ``least Higgs-boson-like'' di-jet mass is improved substantially without biasing the ``most Higgs-boson-like'', bettering the separation of signal and background. The 6C-fit also offers separation power through hypothesis testing with two hypotheses, a $\PZ\PH\PH$ or $\PZ\PZ\PH$ final state. Figure~12 in~\cite{ZHHnote} displays this separation power through the $\chi^2_{\PZ\PZ\PH}/\chi^2_{\PZ\PH\PH}$-ratio, and is used as input to a BDT for the event selection in addition to other observables from the 4C- and 6C-fits. The same figure also shows that there is good agreement between full simulation and SGV in terms of kinematic fitting. 

\section{Updating the ZHH analysis}
\label{sec:zhhanalysis}
The 2014 $\PZ\PH\PH$ analysis effort included two di-Higgs decay modes with $\PH\PH\to 4\PQb$~\cite{Durig:2016jrs, Tian:2013} and $\PH\PH\to \PQb\PAQb\PW\PW^*$~\cite{Kurata:2013}. Both where performed using full GEANT4 based simulations of the ILD concept~\cite{ILDIDR, ILDESU, TDR4, abramowicz2025ilddetectorversatiledetector}. At the time, no event-level combination was possible due to the inclusive approach of the $\PH\PH\to 4\PQb$ which optimised the signal selection to include $\PH\PH\to 4\PQb$ and non-$4\PQb$, preventing a proper combination of the $\PH\PH\to 4\PQb$ and $\PH\PH\to \PQb\PAQb\PW\PW^*$ channels. Instead a 20\% relative improvement was assumed to be expected which gave the combined precision of 26.6\% on $\Delta\uplambda_{SM}/\uplambda_{SM}$. A few key differences between the 2014 analysis and the ongoing work are hightlighted in the following. Firstly, the ongoing analysis is using fast SGV based simulations of the ILD concept for now until a full SM production at 550\,GeV is possible on a longer timescale. Secondly, the signal selection targets $\PH\PH\to 4\PQb$ only for an eventual combination of channels, with the same channels of the $\PZ$-decay as in~\cite{Durig:2016jrs} with $\Pep\Pem\PH\PH$, $\PGmp\PGmm\PH\PH$, $\PGn\PAGn\PH\PH$, $\PQb\PAQb \PH\PH$, and $\PQq\PAQq \PH\PH$ but splitting the neutrino channel by differentiating between $\PZ\PH\PH$ and $\PW\PW$-fusion events as explained below. Thirdly, the set of backgrounds have been expanded since the 2014 analysis which only considered $\PZ\PZ\PH$, $\PZ\PZ\PZ$, $\PQt\PAQt$, $\PW\PW\PZ$, $\PQt\PAQt\PZ$, $\PQt\PAQt\PH$ and 4-fermion backgrounds with two or more $\PQb$-quarks. The ongoing analysis considers the full SM 4-fermion and 6-fermion backgrounds in addition to the specific 8-fermion backgrounds of $\PQt\PAQt\PZ$ and $\PQt\PAQt\PH$. 

The ongoing $\PZ\PH\PH$ analysis is taking shape with first results in the lepton channels being ready. Having implemented the improvements in analysis and reconstruction tools, the analysis strategy is summarised in the following. 
For the event reconstruction, the electrons and muons of are identified through isolated lepton tagging based on BDTs~\cite{Durig:2016jrs}. In case of more than two isolated leptons, the two leptons best complying with a Z-boson are paired, and the rest of the event is clustered into a four jets with the Durham algorithm using the LCFIPlus framework. Then ErrorFlow and the neutrino correction are applied before performing the 4C and 6C fits on the lepton channel signature, as well as tagging the jet flavors with PartT. For the event selection, the preselection is taken from the 2014 analysis and lightly modified to increase the signal training statistics. The event selection then consists of four BDTs trained against the four main backgrounds, $\Pl\Pl\PQb\PQb$, $\Pl\PGn \PQb\PQb\PQq\PQq$, $\Pl\Pl\PQq\PQq\PQq\PQq$, and $\Pl\Pl\PQq\PQq\PH$. The significances of the electron and muon channels are summarised in Table~\ref{tab:llbbbb}, where they are also compared with the results of the extrapolation (see Sec.~\ref{sec:extrapolations}). This comparison shows that not only can we confirm the results of the extrapolation in the lepton channels, we also exceed the projections of the extrapolation in these channels, verifying that the assumptions made were conservative. More details can be found in~\cite{ZHHnote}. The results of Table~\ref{tab:llbbbb} are still a work in progress as they are only based on a first attempt to optimize the event selection with more opportunities to optimise it further.

The re-analysis in the neutrino and hadron channels are also underway, and the neutrino channel will only be briefly commented on here. In the neutrino channel both di-Higgs strahlung and $\PW\PW$-fusion contribute to the same final state of $\PGn\PAGn\PH\PH$. These two neutrino channels have very different interference patterns. The sensitivity to $\uplambda_{\PH\PH\PH}$ can be enhanced by correctly identifying which of the two processes generated the $\PGn\PAGn\PH\PH$ final state. As a proof-of-principle a BDT was trained to separate $\PZ\PH\PH$ events from $\PW\PW$-fusion events using five observables; missing transverse momentum, missing mass, visible energy, visible mass and thrust. At generator level, the two processes in the $\PGn\PAGn\PH\PH$ events were seperated to a good approximation by using a cut on the invariant mass of the generated neutrino pair. The output of the BDT can be found Fig.~19 of~\cite{ZHHnote}, and was used in the extrapolation as well splitting the $\PGn\PAGn\PH\PH$ channel into a $\PW\PW$-fusion contribution and $\PZ\PH\PH$ contribution.  

\begin{table}
      \centering
      \begin{tabular}{ rcc }
        \hline
        $S/\sqrt{S+B}$ & Analysis      & Extrapolation \\ \hline
        $\Pep\Pem\PQb\PAQb\PQb\PAQb$         & \textbf{1.28} & 1.15 \\
        $\PGmp\PGmm\PQb\PAQb\PQb\PAQb$    & \textbf{1.39} & 1.14 \\ \hline
      \end{tabular}
      \caption{Significance of the lepton channels comparing the ongoing analysis and extrapolations~\cite{ZHHnote}.}
      \label{tab:llbbbb}
\end{table}

\begin{figure}[htbp]
    \centering
    \begin{subfigure}{.5\textwidth}
    \centering
        \includegraphics[width=0.95\textwidth]{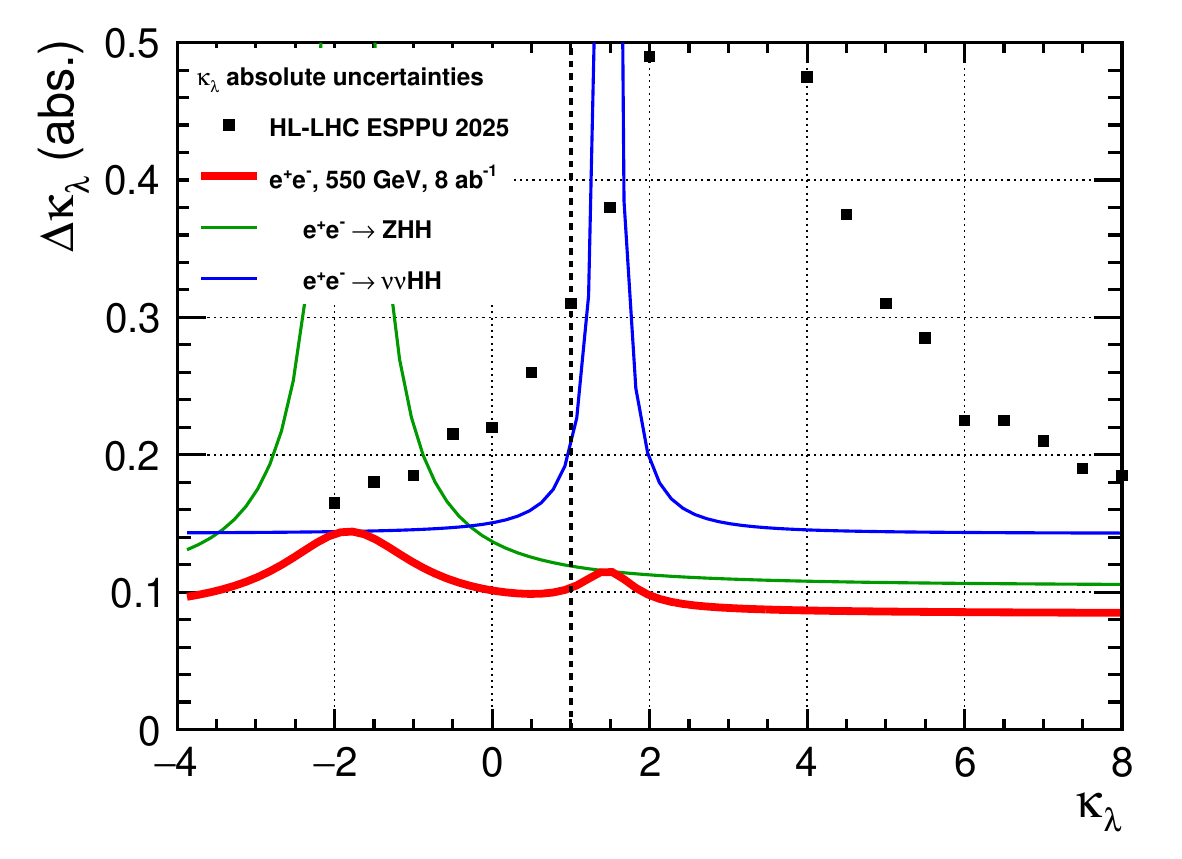}
        \caption{}
        \label{fig:absvskala:550}
    \end{subfigure}\hfill%
    \begin{subfigure}{.5\textwidth}
        \centering
        \includegraphics[width=0.95\textwidth]{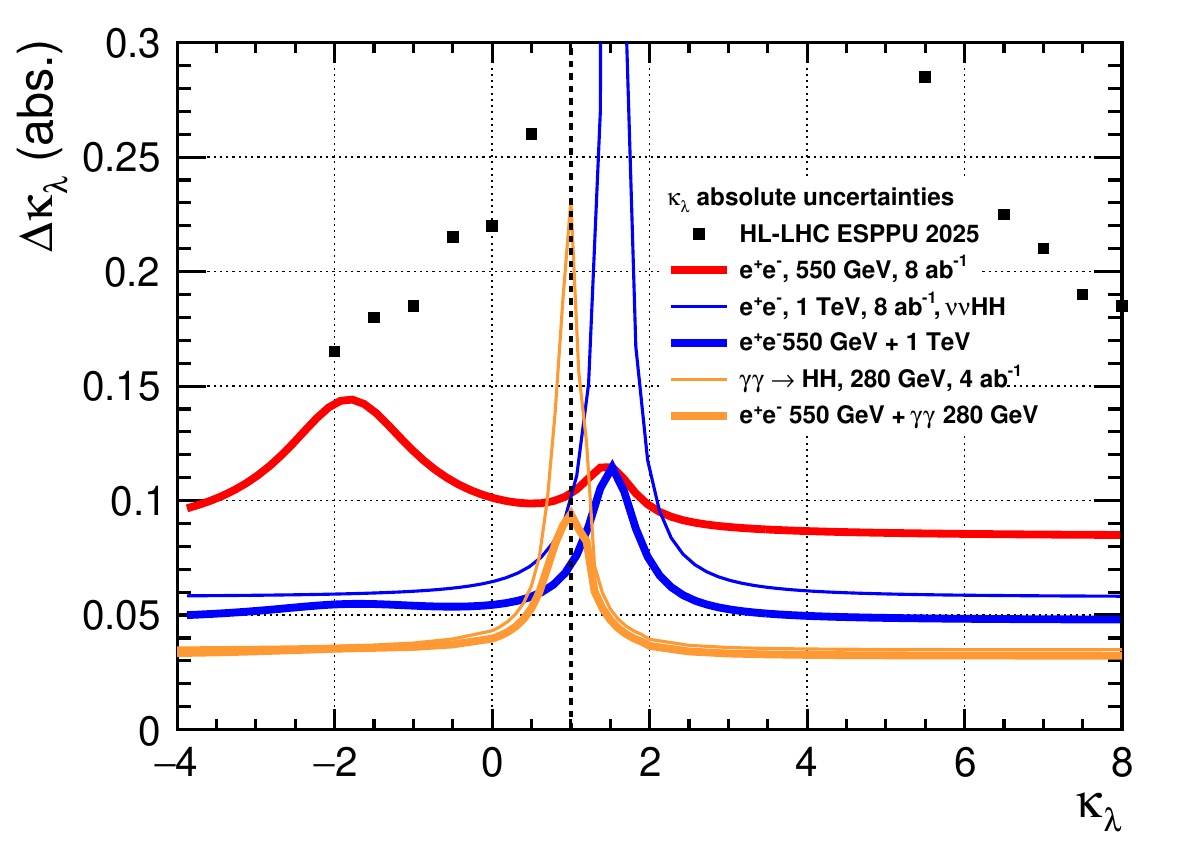}
        \caption{}
        \label{fig:absvskala:beyond}
    \end{subfigure}%
    \caption{Absolute uncertainty on $\kappa_{\uplambda}$ versus itself at LCF550 (a) and beyond LCF550 (b)~\cite{ZHHnote}. }
    \label{fig:absvskala}
\end{figure}

\section{Extrapolations for projections on the SM Higgs self-coupling} \label{sec:extrapolations}
Two key observations of the improvements in flavor tagging and kinematic reconstruction have gone into the extrapolation. As mentioned above, a 10\% efficiency increase in $\PQb$-tagging is assumed for the same level of background rejection which counts to the third power since it is the presence of a third $\PQb$-jet already that separates the signal with $\PH\PH\to4\PQb$ from $2\PQb$-jet backgrounds. From the improvements in kinematic reconstruction, a 10\% increase in event selection effeciency is assumed at the same level of background rejection. The number of signal events, $S$, and background events, $B$, of the 2014 $\PH\PH\to4\PQb$ analysis are taken where, for the time being, the significance is calculated from the simplified measure of $S/\sqrt{S+B}$ (instead of a likelihood ratio). Then $S$ is scaled up according to the expected improvements while $B$ is kept the same for each of the channel of $\Pep\Pem\PH\PH$, $\PGmp\PGmm\PH\PH$, $\PGn\PAGn\PH\PH$, $\PQb\PAQb \PH\PH$, and $\PQq\PAQq \PH\PH$ before being combined. The numbers can be found in Table~9 of~\cite{ZHHnote}. The extrapolation leads to an estimate of 16.6\% uncertainty on the cross section, which is to be compared to the 21.1\% uncertainty found from~\cite{Durig:2016jrs}. The precision improves by imposing the same assumptions from the 2014 analysis on the eventual combination with other channels, with a 20\% relative improvement expected from including the $\PH\PH\to \PQb\PAQb\PW\PW^*$ channel, an 8\% relative improvement expected from including $\PZ\to\PGtp\PGtm$, and 10\% relative improvement expected from including additional Higgs decays not yet studied for future linear colliders. From this we find, that with the ILC programme running at a centre-of-mass energy of 500\,GeV with an integrated luminosity of 4\,ab$^{-1}$, the projected precision on the Higgs self-coupling improves (from 27\%) to 18\% i.e. from advancements in reconstruction tools alone, the ILC500 programme has improved its projections on $\uplambda_{SM}$ to the discovery reach.

The precision is expected to improve further with the revision of running scenarios for future linear $\Pep\Pem$ colliders. By increasing the centre-of-mass energy from 500\,GeV to 550\,GeV, the cross section on $\PZ\PH\PH$ events increases by 16\% and, more dramatically, the cross section for $\PGn\PAGn\PH\PH$ events from $\PW\PW$-fusion increases by about a factor of 2. At 550\,GeV for P$(\Pep,\Pem)=(-80\%,+30\%)$, the neutrino channel can be separated into a $\PZ\PH\PH$ part and $\PW\PW$-fusion part following the procedure explained above in Sec.~\ref{sec:zhhanalysis}. From increasing the centre-of-mass energy to 550\,GeV and adding the $\PW\PW$-fusion channel, the precision on the Higgs self-coupling would already improve to 15\%. With the more ambitious programme of the LCF where the luminosity is doubled from 4\,ab$^{-1}$ to 8\,ab$^{-1}$, and the positron polarisation is doubled from 30\% to 60\% (corresponding to increasing the luminosity by another 10\%), the projections on the precision reach on the Higgs self-coupling can be further improved to 11\%. The projections for various running scenarious are summarised in Table~\ref{tab:extrapolations}.

\begin{table}
      \centering
      \begin{tabular}{ lccc|c }
        \hline
        Collider  & $\sqrt{s}$\,[GeV] & $\mathcal{L}$ [ab$^{-1}$] & $|$P$(\Pem,\Pep)|$ & $\Delta\uplambda_{SM}/\uplambda_{SM}$  \\ \hline
        ILC       & 500               & 4                         & (80\%,30\%)        & 18\%\\
        ILC/C$^3$ & 550               & 4                         & (80\%,30\%)        & 15\% \\
        LCF       & 550               & 8                         & (80\%,60\%)        & 11\% \\ \hline
      \end{tabular}
      \caption{Precision on the SM Higgs self-coupling estimated from extrapolations for various running scenarios of future linear $\Pep\Pem$ colliders~\cite{ZHHnote}.}
      \label{tab:extrapolations}
\end{table}

\subsection{Beyond the SM}
The projected precisions in Sec.~\ref{sec:extrapolations}, with an 11\% uncertainty for the LCF at 550\,GeV, are only valid in the case that the value of $\uplambda_{\PH\PH\PH}$ realised in nature is consistent with the SM value. The precision on $\uplambda_{\PH\PH\PH}$ varies strongly with the value of $\uplambda_{\PH\PH\PH}$, as the value affects both the cross sections and the interference between the diagrams with and without the Higgs self-coupling, hence also impacting the value of the sensitivity factor in Eq.~\ref{eq:precision}. The cross section and cross section uncertainty dependence on $\kappa_{\uplambda}=\uplambda_{\PH\PH\PH}/\uplambda_{SM}$ can be found in Fig.~20 in~\cite{ZHHnote} and the same dependence of the sensitity factor can be found in Fig.~21 also in~\cite{ZHHnote}. The value of $\uplambda_{\PH\PH\PH}$ has a very different effect on the two processes of $\PZ\PH\PH$ and $\PW\PW$-fusion resulting in an important complementarity. This complementarity is well-illustrated in Fig.~\ref{fig:absvskala:550} which shows the absolute uncertainty on $\kappa_{\uplambda}$ as a function of $\kappa_{\uplambda}$ for the two processes, and their combined uncertanity. In ranges of $\uplambda_{\PH\PH\PH}$ where the precision is degraded for one process the other process is able to compensate, resulting in absolute uncertainties at the $\sim10-15\%$ level for all values of $\uplambda_{\PH\PH\PH}$. The same figure also shows the projected sensitivity at HL-LHC which does not have this same complementarity due to very different di-Higgs production modes at a hadron collider, and therefore the precision is degraded for $\kappa_{\uplambda}\gtrsim 2$. Linear colliders also offer opportunities to improve the sensitivity to the Higgs self-coupling further by increasing the centre-of-mass energies or replacing the $\Pep\Pem$ collider, with a $\PGg\PGg$ collider pushing the precision down to the $\sim5-10\%$ level as is seen from Fig.~\ref{fig:absvskala:beyond}. It should be noted that the $\PGg\PGg$ collider projections are based on a recent preliminary study of $\PGg\PGg\to\PH\PH\to4\PQb$ in Delphes simulations of the SiD concept~\cite{Barklow:2025privcom}.

\section{Conclusion}
The re-analysis of di-Higgs production within the ILD Concept is ongoing, for the first time at 550\,GeV. The $\PZ\PH\PH$ analysis is profiting from advancements in event reconstruction and analysis algorithms. Our first attempt in the lepton channel confirms and exceeds the extrapolations made for EPPSU. The re-analyses of the neutrino and hadron channels are also underway. The projections on the Higgs self-coupling at a future linear collider at 550\,GeV submitted for the upcoming update of the European Strategy for Particle Physics~\cite{briefingbook}, are 
  \begin{itemize}
  \item ILC550 @ $\mathcal{L}$ = 4\,ab$^{-1}$: $\Delta\uplambda_{SM}/\uplambda_{SM}\to 15\%$
  \item LCF550 @ $\mathcal{L}$ = 8\,ab$^{-1}$: $\Delta\uplambda_{SM}/\uplambda_{SM}\to 11\%$
  \end{itemize}
with $\sim10-15\%$ percent level precision reach across the range of $\uplambda_{\PH\PH\PH}$ for the LCF and with opportunities at linear colliders to improve that further.

\section*{Acknowledgements}
We would like to thank the LCC generator working group and the ILD software working group for providing the simulation and reconstruction tools and producing some of the Monte Carlo samples used in this study.
This work has benefited from computing services provided by the ILC Virtual Organization, supported by the national resource providers of the EGI Federation and the Open Science GRID.
We thankfully acknowledge the support by the 
the Deutsche Forschungsgemeinschaft (DFG, German Research Foundation) under Germany's Excellence Strategy EXC 2121 ``Quantum Universe'' 390833306. \\
\\
The simulated data used in this work is available by contacting the ILD Concept Group (\href{mailto:ild-et@desy.de}{ild-et@desy.de}). 


\printbibliography{}
\end{document}